\documentstyle[preprint,aps]{revtex}

\newcommand{\be}{\begin{equation}}
\newcommand{\ee}{\end{equation}}
\newcommand{\bea}{\begin{eqnarray}}
\newcommand{\eea}{\end{eqnarray}}

\begin{document}
 \preprint{\small CGPG-95/10-8  gr-qc/9511003}
\draft
\title
 {Black hole solutions in 2+1 dimensions}
\author{Viqar Husain\footnote{
\noindent email: husain@phys.psu.edu\ .
Address after Sep. 15 '95 : Center for Gravitational Physics and
Geometry, Department of Physics, The Pennsylvania State University,
University Park, PA 16802-6300, USA.}}
\address
 {Department of Mathematics and Statistics,\\
University of Calgary, Calgary, Alberta, Canada T2N 1N4.}

\maketitle

\begin{abstract}

We give circularly symmetric solutions for null fluid collapse
in 2+1-dimensional Einstein gravity with a cosmological constant.
The fluid pressure $P$ and energy density $\rho$ are related by
$P=k\rho$ $(k\le 1)$. The long time limit of the solutions are
black holes whose horizon structures depend on the value of $k$.
The  $k=1$ solution is the Banados-Teitelboim-Zanelli black hole
metric in the long time static limit, while the $k<1$ solutions
give  other, `hairy' black hole metrics in this limit.

 \end{abstract}
\bigskip

Three dimensional Einstein gravity provides a model for
exploring many conceptual questions that arise in
four dimensional generally covariant theories. Among these
are various issues in quantum gravity and black hole physics.

The study of black holes in three dimensions  became
possible relatively recently when Banados, Teitelboim and Zanelli
(BTZ) discovered a  black hole solution \cite{btz} in
2+1-dimensional  Einstein gravity with a negative cosmological
constant. Since then there has been substantial work on black hole
physics using this solution \cite{rev}. The general BTZ metric
is characterized, like the  four dimensional  Kerr-Newman metric,
by its mass, angular momentum and electric charge, but is
aymptotically anti-deSitter rather than flat.

One of the questions concerning the BTZ solution is whether it can
arise as the long time (static) limit of collapsing distributions of
matter. This is indeed the case. There is for example, a three
dimensional analog \cite{vhv} of the Vaidya metric \cite{vaidya}
 which describes the collapse of a null fluid. The mass of the
resulting BTZ black hole depends on the parameters characterizing
the collapsing fluid. Other solutions that also describe collapse to
the BTZ black hole have been given \cite{mann}.

This raises the question of whether gravitational collapse of
other types of matter also leads, in the asymptotic long time limit,
to the BTZ black hole. This question is related to what `hairs'
 a 2+1 black hole can have.

In this paper we partially address this question. We find  exact
inhomogeneous and non-static spherically symmetric solutions of the
2+1-dimensional Einstein equations for a collapsing null fluid, with
pressure $P$  and energy density $\rho$ related by the equation of
state  $P=k\rho$. We find that the long time limits of the solutions
give black hole metrics more general than the BTZ metric. The BTZ
metric arises as the special case $k=1$, while the metrics for
$k< 1$ in general give black hole solutions with multiple
apparent horizons.

An inverted approach is used to find the solutions. First
the stress-energy tensor is determined from the metric. Then
the equation of state and the dominant energy condition are
{\it imposed} on its eigenvalues.  This leads to an equation for
the metric function, which is easily solved.
\footnote{The same procedure has been used by the author
to find  null fluid collapse solutions in four-dimensional
Einstein gravity \cite{vh4d}.} The precise form of
the stress-energy tensor is then displayed. The forms of
 the apparent horizons in some cases are  discussed.

The general circularly symmetric metric in 2+1 dimensions
may be written as
\be
ds^2  =  -e^{2\psi(r,v)}\ F(r,v)\ dv^2 +
2 e^{\psi(r,v)}\ dvdr + r^2 d\theta^2, \label{metric}
\ee
where $0\le r\le \infty$ is the proper radial coordinate,
$-\infty \le v \le \infty $ is an advanced time
coordinate and $0\le \theta \le 2\pi $ is the angular coordinate.
The mass function $m(r,v)$ is defined by
 $F(r,v) = 1 - 2m(r,v) /r$. The Einstein equations
\be
G_{ab} + \Lambda\ g_{ab} = 2\pi\ T_{ab}
\ee
 give
\bea
 \psi' &=& 2\pi r\ T_{rr}\ , \\
\dot{m} &=& 2\pi r^2\ T_v^{\ r}, \\
{m\over r} - m' &=& -\Lambda\ r^2 + 2\pi r^2\ T_v^{\ v},
\eea
 where the prime and dot denote partial derivatives with respect to
$r$ and $v$.  We will consider the case $\psi(r,v) = 0$, which
means $T_{rr}=0$.

The stress-energy tensor derived from  the above metric
may be diagonalized to give the energy density and the principal
pressures. The eigenvalue problem is $T_a^{\ b}U_b = \lambda U_a$.
The $\theta$  part of the tensor is already diagonal
with pressure eigenvalues
\be
 P\equiv T_\theta^{\ \theta} = {G_{\theta\theta} + \Lambda\ r^2 \over
2\pi r^2} =
{1\over 2\pi r^3}\ ( -r^2\ m'' + 2\ r\ m' - 2\ m + \Lambda\ r^3).
\label{P}
\ee
Since $\psi=T_{rr}=0$ (by choice), $T_v^{\ v} =  T_r^{\ r}$ and
$T_r^{\ v}=0$,  therefore the $v$-$r$ part of the matrix to be
diagonalized is
\be
T_a^{\ b} = \left( \begin{array}{cc}
                    T_v^{\ v} & T_v^{\ r}   \\
                    0 &  T_v^{\ v}
                    \end{array}
             \right),
\ee
This has one eigenvalue $\lambda$ which gives the energy density
$\rho$, namely
\be
 \rho \equiv -\lambda = - T_v^{\ v} = {1\over 2\pi r^3}\
( r\ m' - m - \Lambda\ r^3 ).
\label{rho}
\ee
The corresponding  eigenvector is $v_a=(1,0,0)$, (in the
coordinates $(v,r,\theta)$), and is lightlike. Therefore
the  stress-energy  tensor, (which follows from $\psi=0$), is
of Type II \cite{hawkell}.\footnote{Stress-energy tensors are
classified by their eigenvectors. In four dimensions, Type I
stress-energy tensors have one timelike and three spacelike
eigenvectors, and Type II tensors have one lightlike and two
spacelike eigenvectors. Most physical fields are of Type I,
except for certain null fluids, which are of type Type II.}
 Its non-vanishing components are
\bea
T_{vv} &=& \rho\ (1- {2m\over r}\ ) + {\dot{m}\over 2\pi r^2},
\nonumber \\
 T_{vr} &=& -\rho,  \nonumber \\
T_{\theta\theta} &=& P\ g_{\theta\theta}\ ,
 \eea
which may  be succinctly written using the lightlike vectors
$v_a$ and  $w_a=(F/2,-1,0)$ as
\be
T_{ab} =  {\dot{m}\over 2\pi r^2} \ v_av_b +
\rho\ (v_aw_b + v_bw_a) + P\ (g_{ab} + v_aw_b + v_bw_a)\ .
\label{set}
\ee
The last two terms on the right hand side make up a perfect
fluid-like contribution to the stress-energy tensor, with the
difference that the velocities are lightlike.
A stationary observer with 3-velocity  $S^a=(1/\sqrt{F},0,0)$
and a rotating observer with 3-velocity
$R^a=(\sqrt{2/F},0,1/r)$
see respectively the energy densities
\bea
T_{ab}S^aS^b &=& {\dot{m}\over 2\pi r^2 F} + \rho\ , \\
T_{ab}R^aR^b &=& 2\ (\ {\dot{m}\over 2\pi r^2 F} + \rho\ ) + P\ .
\eea

The stress-energy tensor (\ref{set}) satisfies the dominant energy
condition if the following three conditions are met \cite{hawkell} :
\be
P \ge 0,  \ \ \rho\ge P, \ \ {\rm and} \ \
T_{ab}w^a w^b > 0,
\label{de}
\ee
  To  satisfy the first two of these conditions
(\ref{de}), we {\it impose} the  equation of state
$P=k\rho$, with $k\le 1$.  This gives the equation
 \be
 r^2\ m'' - r\ (2-k)\ m' + (2-k)\ m = \Lambda\ (1+k)\ r^3
\label{masseq}
\ee
for the mass function, which has the general solution
\be
m(r,v) =  g(v)\ r + h(v)\ r^{2-k} + {\Lambda\ r^3 \over 2}\ ,
\label{mfn}
\ee
 where $g(v)$ and $h(v)$ are arbitrary functions. Therefore from
 (\ref{P}) and (\ref{rho}) we have explicitly that
\be
P = k\ {(1-k)\ h(v) \over 2\pi\ r^{k+1} } = k \rho.
\ee
Hence we must have  $h(v)\ge 0$ for positive pressure and
energy density. Note that when $k=1$, $P=\rho=0$, and as we
 will see below, this gives the BTZ metric in the long time
limit.

The last requirement in (\ref{de}) for the dominant energy
condition leads  to
\be
\dot{m} =  \dot{g}(v)\ r + \dot{h}(v)\ r^{2-k} > 0.
\ee
Physically this means that the matter within a radius $r$ increases
with time, which corresponds to an implosion.  This is easily
satisfied by choosing $\dot{g}(v)$ and $\dot{h}(v)$ to be positive
everywhere.

In summary, we have shown that the metric
\be
ds^2 = -(1 - 2\ g(v) - 2\ h(v)\ r^{1-k} -\Lambda\ r^2 )\ dv^2
+ 2\ dv dr + r^2 d\theta^2 \label{sol1}
\ee
 is a solution of the 2+1-dimensional Einstein equations for the null
fluid  stress-energy tensor (\ref{set}) with $P=k\rho$.
Since $k\le 1 $, the cosmological constant must be negative for
this metric to be asymptotically Lorentzian, in which case it is
asymptotically anti-deSitter. Therefore in the following we
set $\Lambda = -1/l^2$.

There are two special cases of this solution which are already known.
One is  the  2+1 dimensional analog \cite{vhv} of the Vaidya metric
 \cite{vaidya}, which arises for $k=1$.  This means vanishing $\rho$
and $P$. Then the only non-vanishing component of the stress-energy
tensor (\ref{set}) is $T_{vv} = \dot{m}/2\pi r^2 = \dot{g}(v)/2\pi r$.
The other is the BTZ black hole\cite{btz}, which arises when
$g(v)$=constant, in addition to  $k=1$.
(We note that when $k=1$, $h(v)$ disappears from the stress-energy
tensor (\ref{set}),  therefore setting $k=1$ is the same as
setting  $h(v)=0$).

 We now consider the evolution of the apparent horizons
for some cases of the metric (\ref{sol1}). The
radius of the apparent horizon $r_{AH}$ is given by the
equation  $g^{ab}\partial_a r\partial_b r = 0$,
which gives the equation $m(r,v;c_i)=r/2$ for $r_{AH}(v;c_i)$.
The $c_i$ are any constants that appear in the mass function.
The long time $v\rightarrow\infty$ limit of $r_{AH}$
gives   the asymptotic radius of the apparent horizon as a function
of the constants $c_i$. This is a measure of the black hole mass
for asymptotically flat or (anti)-deSitter spacetimes, namely
$M_{BH}:= \lim_{v\rightarrow\infty} r_{AH}(v;c)/2$.

We focus on the cases where the  $v\rightarrow -\infty$ limit
of  $g(v)$ and $h(v)$ are zero, so that there are no initial
horizons, and their $v\rightarrow +\infty$ limit are finite
non-zero constants, so that there are final horizons. To be
specific,  we assume in  the  following  that
$\lim_{v\rightarrow \infty}g(v) = A\ (>0)$, and
$\lim_{v\rightarrow \infty} h(v) = B\ (>0)$. These conditions may
be met easily by appropriate choices of the arbitrary functions
$g(v)$ and $h(v)$ in the solutions.
 The radii of the apparent horizons in the $v\rightarrow \infty$
limit of the metric (\ref{sol1}) are then given  by
\be
  r^2  - 2 B\ l^2\ r^{1-k} + l^2\ ( 1 - 2A )  = 0\ . \label{ah1}
\ee
This equation leads to a range of polynomial equations for the
asymptotic apparent horizons depending on the values of $k\le 1$.
We will consider a few special cases.

For $k=1$ the radius of the apparent horizon is
\be
r_{AH} = l\  \sqrt{2A - 1},
\ee
which is the single horizon of the circularly symmetric
uncharged BTZ black hole. (As noted above, $k=1$ also implies
$B=0$ ).

For $k=0$, the pressure vanishes but the energy density
$\rho$ does not. The apparent horizon equation (\ref{ah1})
is quadratic with solutions
\be
r_{AH} = l^2\ B   \pm   \sqrt{ l^4\ B^2 - l^2\ (1-2A ) } \ .
\label{k=0}
\ee
This is like the Reissner-Nordstrom black hole in four
dimensions with charge $l\ \sqrt{1-2A}$ and mass $l^2\ B$, with the
extreme case occuring when $ l\ B = \sqrt{1-2A}$. An important
difference however is that there is no curvature singularity at
$r=0$ in the metrics (\ref{sol1}). We note that the
charged BTZ black hole does not give Eqn. (\ref{k=0}) for the
horizons \cite{vhv} because, among other things, the electric
potential in two spatial dimensions is proportional to $\ln r$.

The next simplest case is $k=1/2$ for which  Eqn. (\ref{ah1})
is quartic. All other values of $k$  give higher
order polynomials which may be solved numerically.  The main
observation is that in general there will be multiple horizons,
and the long time limits will give black hole solutions more general
then the BTZ black hole. Therefore 2+1-dimensional black holes
can have `null fluid hair' in the static limit.

An explicit example of a hairy black hole is provided by taking the
arbitrary functions $g(v)$ and $h(v)$ in the mass function (\ref{mfn})
to be $g(v)=A\tanh v$ and $h(v)= B\ (1+ \tanh v)$, for positive
constants $A$ and $B$. (The dominant energy condition is satisfied
for these choices). Then in the $v\rightarrow\infty$ static limit,
the mass function becomes  $m(r)= A\ r + 2B\ r^{2-k} + \Lambda\ r^3/2$.
Therefore the constant $A$ determines the black hole mass, and $B$
represents the `null fluid hair' of the resulting static black hole.
There are  of course  many possible choices for $g(v)$ and $h(v)$
that lead to such  black holes in the static long time limit.

In conclusion, we have given a new class of null fluid
collapse solutions (\ref{sol1}) of the 2+1 dimensional Einstein
equations with negative cosmological constant for the stress-energy
tensor (\ref{set}). A special case of these solutions is
the BTZ black hole.  It would be of interest to study the  global
structures and thermodynamics of these black hole metrics.
\bigskip

This work was supported by the Natural Science and Engineering
Research Council of Canada.

\end{document}